%% file: main.tex
\documentclass[10pt,twocolumn,letterpaper]{article} 
\usepackage{cvpr}  
\input{preamble} 
\definecolor{cvprblue}{rgb}{0.21,0.49,0.74}
\usepackage[pagebackref,breaklinks,colorlinks,allcolors=cvprblue]{hyperref}
 
\def\paperID{*****}  
\def\confName{CVPR}
\def\confYear{2025} 
\title{EmoVLM-KD: Fusing Distilled Expertise with Vision-Language Models for Visual Emotion Analysis} 
\author{
SangEun Lee$^{1}$, Yubeen Lee$^{2}$, Eunil Park$^{2,}\thanks{Corresponding Author}$ \\
$^{1}$Electronics and Telecommunications Research Institute \\
$^{2}$Sungkyunkwan University \\
{\tt\small sange1104@etri.re.kr, lybin1070@gmail.com, eunilpark@skku.edu
}}

\begin{document}
\maketitle
\input{sec/abstract}     
\input{sec/main} 
{
    \small
    \bibliographystyle{ieeenat_fullname}
    \bibliography{main}
} 

\end{document}

%% file: preamble.tex
\usepackage{multirow}
\usepackage{colortbl}
\usepackage{xcolor} 
\usepackage{tabularx}
\usepackage{float}
\usepackage[accsupp]{axessibility}

%% file: sec/abstract.tex
\begin{abstract}
Visual emotion analysis, which has gained considerable attention in the field of affective computing, aims to predict the dominant emotions conveyed by an image. Despite advancements in visual emotion analysis with the emergence of vision-language models, we observed that instruction-tuned vision-language models and conventional vision models exhibit complementary strengths in visual emotion analysis, as vision-language models excel in certain cases, whereas vision models perform better in others. This finding highlights the need to integrate these capabilities to enhance the performance of visual emotion analysis. To bridge this gap, we propose EmoVLM-KD, an instruction-tuned vision-language model augmented with a lightweight module distilled from conventional vision models. Instead of deploying both models simultaneously, which incurs high computational costs, we transfer the predictive patterns of a conventional vision model into the vision-language model using a knowledge distillation framework. Our approach first fine-tunes a vision-language model on emotion-specific instruction data and then attaches a distilled module to its visual encoder while keeping the vision-language model frozen. Predictions from the vision language model and the distillation module are effectively balanced by a gate module, which subsequently generates the final outcome. Extensive experiments show that EmoVLM-KD achieves state-of-the-art performance on multiple visual emotion analysis benchmark datasets, outperforming the existing methods while maintaining computational efficiency. The code is available in \url{https://github.com/sange1104/EmoVLM-KD}.
 \end{abstract}

%% file: sec/main.tex
\section{Introduction}
\label{sec:intro}

\begin{figure}[ht]
    \centering
    \includegraphics[width=0.5\textwidth]{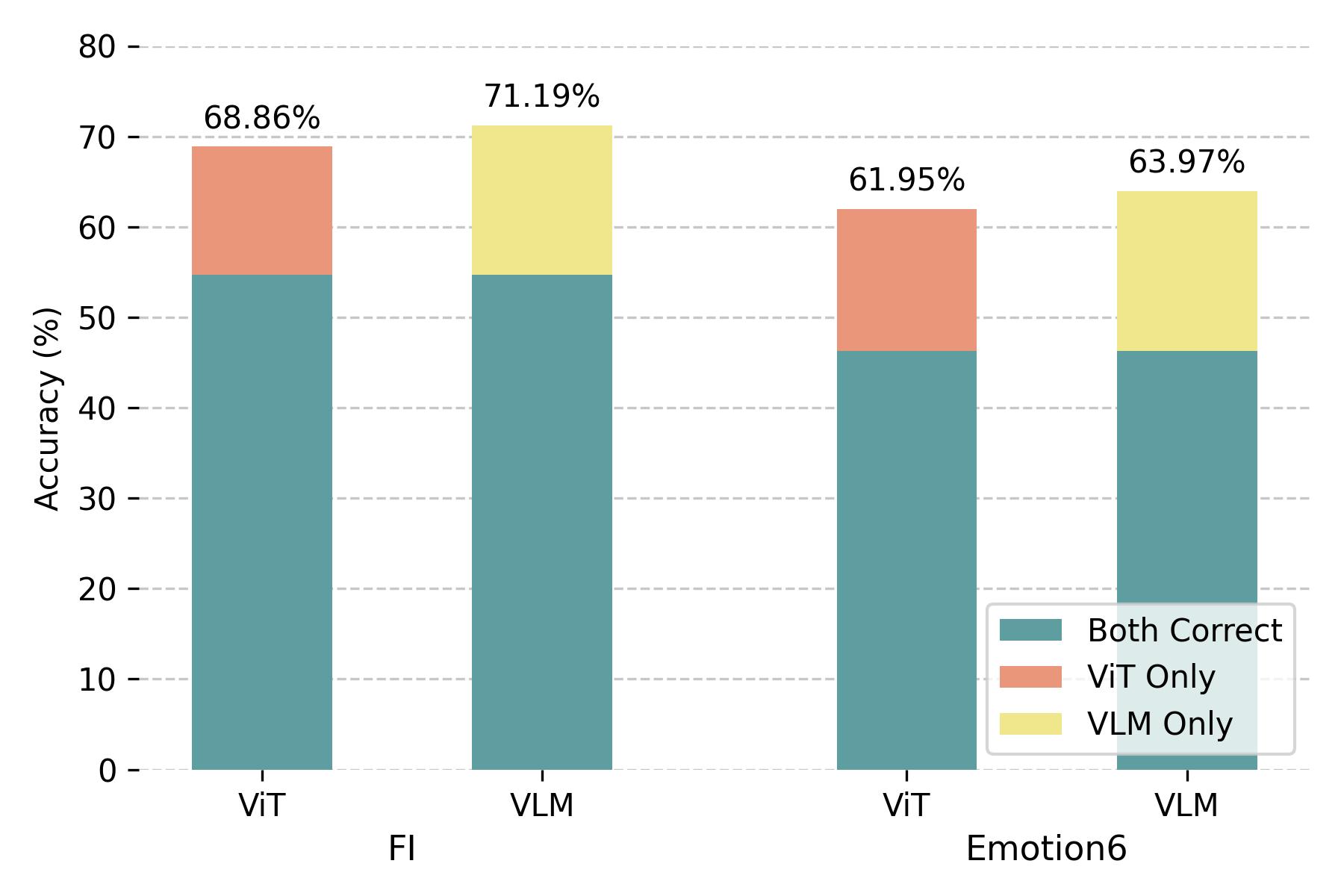}
    \caption{Performance comparison between Vision Transformer (ViT) and VLM. ViT- and VLM-only represent the proportion of instances correctly predicted by each model individually, whereas the other model fails. Although the overall performance of the two models is similar, the significant proportions of ViT-only and VLM-only suggest the need for the two models to complement each other.} 
    \label{fig:introduction}
\end{figure} 

Visual Emotion Analysis (VEA) is a crucial task in the field of computer vision, focusing on predicting the dominant emotion category conveyed by a given image~\cite{wang2022systematic}. Unlike traditional vision tasks that focus solely on recognizing objective information (e.g., semantic segmentation~\cite{strudel2021segmenter} and scene recognition~\cite{herranz2016scene}), VEA involves capturing subjective and ambiguous emotional information, making it significantly more challenging~\cite{yang2018retrieving}. By facilitating the understanding of emotions, VEA provides a foundation for improving human-machine interactions and enables broad applications in domains such as mental health~\cite{fei2020deep}, social relation inference~\cite{fang2015relational}, and advertising~\cite{mitchell1986effect, shukla2020recognition}.  

With recent advancements in Large Language Models (LLMs), there has been growing interest in leveraging their power for multimodal tasks, including VEA~\cite{takahashi2024personalized, cheng2024emotion}. Specifically, EmoVIT~\cite{xie2024emovit} introduced an instruction tuning method to VEA by incorporating emotion-specific instruction data with a Vision-Language Model (VLM), thereby significantly enhancing emotion prediction performance. This demonstrates that VLMs can achieve substantial progress in VEA, and even surpass traditional approaches that do not rely on LLMs. 

On the other hand, we observed that instruction-tuned VLMs and conventional vision models, which are typically represented by Convolutional Neural Network (CNN)-based architectures (e.g., VGG~\cite{simonyan2015very} and ResNet~\cite{he2016deep}) and transformer-based vision models~\cite{dosovitskiy2020image}, have fundamentally different approaches to VEA owing to their reliance on different sources of information. In particular, VLMs leverage their extensive pre-trained linguistic knowledge to analyze emotions, whereas these vision models depend solely on visual features, resulting in inconsistent and often contradictory predictions. That is, as shown in Figure~\ref{fig:introduction}, some cases are correctly predicted by VLMs but not by vision models, whereas others are the opposite. This implies that each model has its own strengths and significant weaknesses, making it difficult to fully understand the emotions in images. To overcome this limitation, our work focuses on combining their strengths to boost the VEA performance.
 
In this work, we propose EmoVLM-KD, a novel instruction-tuned VLM for VEA with a lightweight module that is distilled from conventional vision models. Although an ensemble of VLMs and conventional vision models can improve performance, employing two such large models simultaneously incurs prohibitively high computational costs. To address this issue, we aim to transfer knowledge from conventional vision models to a VLM using only 0.003\% of their parameters.

Specifically, a VLM is first instruction-tuned using emotion-specific instruction data to learn how humans interpret emotions in images. Here, we adopted the instruction data collection strategy proposed in EmoVIT, which employs GPT-4 to automatically generate detailed instructions for images related to emotional information. In contrast, a conventional vision model is trained on the VEA task in a domain-specific manner. Subsequently, a projection module is attached to the visual encoder of the VLM. When the VLM remains frozen, this additional module is trained via knowledge distillation to mimic the prediction patterns of the conventional vision model. Finally, a gating mechanism is introduced to dynamically balance the contributions of the distillation module and VLM, allowing the model to adapt its final prediction based on both signals.

The contributions are summarized as follows:
\begin{itemize}
    \item Based on the finding that VLMs and conventional vision models exhibit complementary strengths in VEA, we introduce a novel approach to transferring knowledge from domain-specific vision models into VLMs.
    \item To mitigate the high computational cost of employing two large models simultaneously,  we adopt a knowledge distillation framework to inject the vision model’s expertise into the instruction-tuned VLM, achieving this with only a minimal increase in parameters for training.
    \item  Extensive experiments demonstrate that our approach outperforms existing VEA approaches and achieves state-of-the-art results across multiple VEA datasets.  
\end{itemize} 

\section{Related Work}
\label{sec:formatting} 
VEA is the task of understanding and predicting subtle emotions embedded in an image based on visual elements. Early studies primarily relied on low-level visual features such as color, texture, and shape, developing emotion prediction models based on handcrafted features ~\citep{machajdik2010affective, siersdorfer2010analyzing, lu2012shape}. However, these methods have limitations in effectively capturing semantic associations with emotions and explaining the process by which emotions are formed in complex scenes. To address these shortcomings, ~\citet{rao2016multi} employed multiple instance learning and multi-scale blocks to analyze regional emotions within an image, and leveraged probabilistic latent semantic analysis to predict emotions using mid-level representations. Furthermore, large-scale ontology construction for sentiment analysis ~\citep{borth2013sentibank} and studies of visual characteristics that evoke human emotions in complex scenes ~\citep{lu2017investigation} have demonstrated the potential of mid-level representations for emotion prediction.
 
With the advancement of deep learning technologies, high-level feature extraction methods have been introduced, where CNN and Transformer-based models have significantly improved emotion prediction performance by learning not only detailed features of images but also their global features ~\citep{yang2018weakly, he2019multi, zhao2019pdanet, yang2021stimuli, xu2022mdan}. Building on this, ~\citet{you2017visual} proposed a method that utilizes an attention mechanism to automatically identify emotional regions in images. Additionally, ~\citet{lee2022osanet} introduced the Object Semantic Attention Network (OSANet), which leverages the semantic information of objects, while ~\citet{cen2024masanet} proposed a Multi-Aspect Semantic Auxiliary Network (MASANet), which utilizes cross-modal generation to expand modality representation and introduces cross-modal gating and adaptive fusion modules to enhance visual sentiment analysis performance.

\begin{figure*}[ht]
    \centering
    \includegraphics[width=\textwidth]{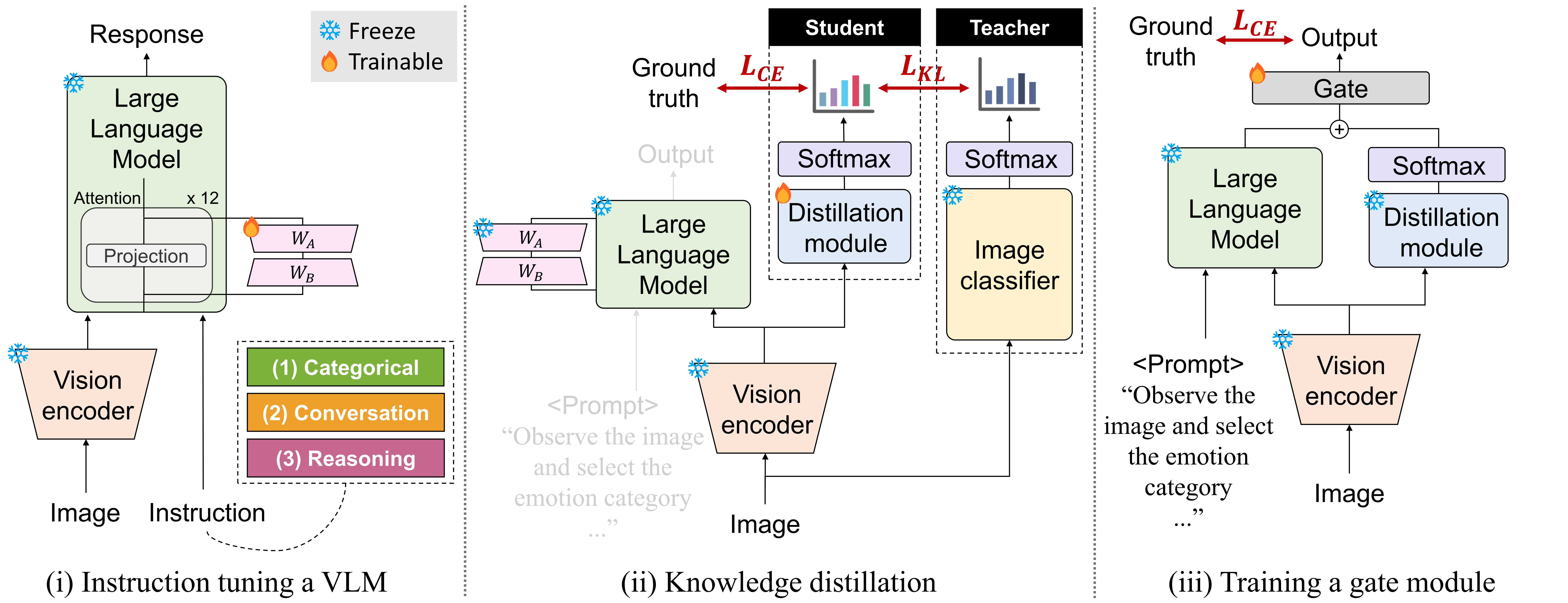}
    \caption{The overall framework of EmoVLM-KD is trained in three stages. (i) Stage 1: a VLM is trained using three types of emotion instruction data. (ii) Stage 2: ViT serves as the teacher, while the distillation module added to the visual encoder acts as the student, distilling the teacher’s knowledge. (iii) Stage 3: The predictions from the LLM and the distillation module are integrated through a gating mechanism to derive the final output.} 
    \label{fig:model_architecture}
\end{figure*} 
 
Recently, approaches utilizing vision–language models (VLMs) have been actively studied. ~\citet{xie2024emovit} enhanced emotion inference and classification performance using the EmoVIT model, which leverages Visual Instruction Tuning (VIT). Additionally, ~\citet{cheng2024emotion} proposed the emotion-LLaMA model for multimodal emotion recognition, presenting a novel approach that integrates audio, visual, and textual inputs. In this study, we focus on the phenomenon in which VLMs and well-established vision models produce either consistent or divergent predictions using the same data. To address this, we aim to achieve more efficient and accurate visual emotion recognition by employing knowledge distillation to transfer the knowledge of well-established vision models into VLMs.
 
\section{Method}
\subsection{Overview}
As shown in Figure~\ref{fig:model_architecture}, the overall framework of EmoVLM-KD is trained sequentially through three key stages: (i) instruction tuning a VLM, (ii) knowledge distillation stage that transfers knowledge from a domain-specific vision model, and (iii) training a gate module that integrates the outputs from the instruction-tuned VLM and distillation module to produce the final prediction. 

\subsection{Instruction Tuning VLM}
To adapt large VLMs for specific purposes, several fine-tuning approaches were presented, such as instruction tuning~\cite{liu2023visual}, prompt tuning~\cite{lester2021power}, and prefix tuning~\cite{li2021prefix}. In this study, we follow the instruction tuning approach by creating emotion-specific instruction datasets for VLMs, as suggested by EmoVIT~\cite{xie2024emovit}.

For instruction tuning of images annotated with emotion labels, GPT-4 was employed to generate detailed responses describing their visual content based on the given instructions. This process facilitates the automatic generation of (image, instruction, and response) triplets. The resulting instruction datasets are categorized into three types: categorical, conversation, and reasoning. 

\begin{itemize}
    \item \textbf{Categorical} instructions prompt the model to select the most appropriate emotion category for a given image, with the response of a specific emotion label. These instructions provide emotion categories as selectable options that vary according to the defined emotion categories in each VEA dataset. Unlike other instruction types, categorical instructions are composed of fixed prompts paired with predefined emotion labels annotated on images, serving as a fundamental method for evaluating the model performance in emotion prediction. An example of categorical instructions is as follows. \\ \\
    {\small \texttt{\textbf{Question}: Observe the image and select the emotion category that best matches this image from the following 8 categories: amusement, anger, awe, contentment, disgust, excitement, fear, and sadness. Answer in dictionary form as follows: \{'emotion':'amusement'\} } \\
    \texttt{\textbf{Response}: \{'emotion': 'contentment'\}}}\\

    \item \textbf{Conversation} instructions are further divided into two subtypes: basic interaction and advanced interaction. Basic interaction involves question-answer pairs that describe key elements of the image, whereas advanced interaction includes more complex and philosophical question-answer pairs. In our work, we only employ the basic interaction type, as follows. \\ \\
    {\small \texttt{\textbf{Question}: Observe the image and describe key elements of the image.} \\
    \texttt{\textbf{Response}: The image features a golden retriever with soft, relaxed fur, lying on green grass. The dog's expression appears calm, with slightly drooping eyelids and a gentle posture. The natural setting contributes to an overall sense of warmth and serene familiarity.}}\\
    \item \textbf{Reasoning} instructions are designed to prompt the model to infer the emotion conveyed by the image through a question-answer format. In addition to the various logical reasoning examples used in EmoVIT, we include a discourse that explores the complex relationships between images and annotated emotion labels, as follows. \\ \\
    {\small \texttt{\textbf{Question}: Observe the image and describe the process of inferring the emotions conveyed in the image.} \\
    \texttt{\textbf{Response}: The dog's relaxed and calm demeanor indicates a state of peace and comfort. Its soft fur and gentle expression evoke feelings of warmth and affection. The lush green grass enhances the peaceful nature of the scene.}}\\

To train a VLM, we adopt the QLoRA technique~\cite{dettmers2023qlora}, focusing specifically on the query, key, and value projection layers within the attention mechanism of the language model. This method allows the model to learn emotion-specific tasks without disrupting the pre-existing knowledge of the VLM. Through this approach, the model effectively integrates emotion-specific instructions while preserving its foundational vision-language capabilities.
\end{itemize}

\subsection{Knowledge Distillation}
The use of knowledge distillation in this study is motivated by the substantial computational memory required to simultaneously run large-scale VLMs and domain-specific vision models. Notably, VLMs inherently contain a visual encoder module pre-trained on extensive image datasets, enabling them to capture general visual knowledge. This study aims to effectively transfer the reasoning patterns of conventional vision models by introducing shallow layers on top of the visual encoder within the VLM. By leveraging VLM's general visual knowledge, this approach facilitates efficient adaptation for emotion category prediction from a given image.

Specifically, we employ a pre-trained vision model as the teacher model, which predicts an emotion category by generating a probability distribution over fixed emotion classes and selecting the class with the highest probability. During training, the distillation module serves as the student learning to replicate the prediction patterns of the teacher model. 

In line with traditional knowledge distillation methods~\cite{zhang2018deep, tiancontrastive, guo2020online}, KL divergence loss~\cite{hinton2015distilling} was employed to enable the student model to capture the predictive pattern of the teacher model, simultaneously incorporating cross-entropy loss based on ground-truth labels to enhance classification accuracy.

First, the KL divergence loss $L_{KD}$ aims to minimize the difference between the probability distributions predicted by the teacher and student models, defined as follows:

\begin{equation}
    L_{\text{KD}} = \tau^2 \sum_{i} p_t^\tau(i) \log\frac{p_t^\tau(i)}{p_s^\tau(i)}
\end{equation}
where $p_t^\tau$ and $p_s^\tau$ denote the softened probability distributions of the teacher and student models, respectively. These distributions are softened using the temperature parameter 
$\tau$, which controls the confidence level in the model's predictions. A higher $\tau$ value makes predictions less confident, creating a smoother distribution that allows the student model to learn more effectively from the teacher. In addition, $i$ refers to the index of the predicted classes (i.e., emotion categories).

In contrast, cross-entropy loss $L_{CE}$~\cite{mao2023cross} calculates the difference between the true label distribution and the predicted probability distribution by penalizing predictions that deviate from the true labels, defined as follows:

\begin{equation}
    L_{\text{CE}} = -\sum_{i} y_i \log p_s(i)
\end{equation}
where $y_i$ is a one-hot encoded value indicating the true label for the $i_{th}$ emotion category and $p_s(i)$ represents the predicted probability for the $i_{th}$ emotion category by the student model.

The overall loss function $L_{total}$ is a weighted combination of the following two losses.

\begin{equation}
    L_{total} = \alpha L_{\text{KD}} + (1 - \alpha) L_{\text{CE}}
\end{equation}
where $\alpha$ denotes a weighting factor that balances the contributions of the two losses. While training, we set the alpha value to 0.5. In Section~\ref{sec:ablation_3}, we conducted an ablation study to determine the optimal alpha value.

At this stage, only the additional distillation module is trained, whereas the rest of the model, including the visual encoder, is kept frozen. This strategy ensures that the distillation module learns to predict emotions by interpreting the general visual knowledge embedded in the frozen visual encoder, thereby aligning it with the emotion-relevant information encoded by the teacher model.

\subsection{Gate Module}
Finally, our framework operates as follows: given an input image, an image and the corresponding text prompt instructing the model to predict the emotion are passed to the visual encoder and LLM, respectively. The visual encoder produces image features that are subsequently forwarded to both LLM and distillation modules.

LLM directly predicts an emotion category, which is transformed into a one-hot encoded vector. The distillation module predicts the probability distribution over emotion categories. These two outputs were concatenated and passed into the gate module.

The gate module integrates the concatenated output to predict the probability distribution across all the emotion categories. Specifically, the concatenated vector is passed through a linear layer that maps it to the final dimension corresponding to the number of emotion categories. This transformation can be represented as follows:

\begin{equation}
    \hat{y} = W \cdot h + b
\end{equation}
where $h$ is the concatenated vector, $W$ is the weight matrix of the linear layer, $b$ is the bias term, and $\hat{y}$ represents the final output logits over the emotion categories. In Section~\ref{sec:ablation_4}, we describe an ablation study conducted to explore the optimal structure of the gate module.

The final prediction is optimized using cross-entropy loss with the ground truth emotion labels, enabling the model to learn to balance the contributions of the LLM and distillation module effectively. Also, while the visual encoder and distillation module remain frozen, only the gate module is trained.

\section{Experimental Results}
\subsection{Datasets}
To train and evaluate the proposed framework, we utilized five benchmark datasets that are widely adopted in VEA. These datasets were labeled with different emotion hierarchies, allowing us to assess the robustness and adaptability of the proposed model across various standards of emotion classification.

First, we employed two datasets annotated based on Mikel's eight basic emotion categories (i.e., amusement, anger, awe, contentment, disgust, excitement, fear, and sadness)~\cite{mikels2005emotional}: EmoSet~\cite{yang2023emoset}, which contained 118,102 images, FI~\cite{you2016building} with 21,824 images. These datasets provide a comprehensive set of images labeled with one of eight emotion categories, enabling a robust evaluation of the model's ability to recognize a diverse range of emotions.  

In addition, we included the Emotion6 dataset~\cite{peng2015mixed}, which comprises 1,980 images annotated according to Ekman's six basic emotions (i.e., anger, surprise, disgust, joy, fear, and sadness)~\cite{ekman1999basic}.

Furthermore, we used the Flickr and Instagram datasets~\cite{katsurai2016image}, which consist of 60,738 and 42,832 images, respectively. These datasets were annotated with binary emotion labels—positive or negative—offering a simplified yet distinct perspective for evaluating the model's performance. 

\begin{table*}[t]
    \centering
    \renewcommand{\arraystretch}{1}  
    \setlength{\tabcolsep}{5pt} 
    \begin{tabularx}{\linewidth}{>{\raggedright\arraybackslash}m{4cm} >{\centering\arraybackslash}X >{\centering\arraybackslash}X >{\centering\arraybackslash}X >{\centering\arraybackslash}X >{\centering\arraybackslash}X}  \hline 
        
        \textbf{Model}&  \textbf{Emoset}&  \textbf{FI}&  \textbf{Emotion6}&  \textbf{Flickr}& \textbf{Instagram}\\\hline
 \multicolumn{6}{c}{\textbf{Conventional Vision Models}}\\ 
         VGG~\cite{simonyan2015very}&  72.27&  51.21& 37.21&  80.52& 77.19\\  
         ResNet~\cite{he2016deep}&  74.04&  64.74&  54.99& 82.73 & 81.45\\  
         ViT~\cite{dosovitskiy2020image}& 78.27& 68.86& 61.95& 85.54&85.41\\ \hline 
 \multicolumn{6}{c}{\textbf{VEA-specific Models}}\\ 
         WSCNet~\cite{yang2018weakly}&  76.32&  70.07&    58.47&  81.36& 81.81\\  
         Stimuli-aware~\cite{yang2021stimuli}&  78.40&  72.42& - &  85.64& 84.90\\  
         MDAN~\cite{xu2022mdan}& 75.75& 76.41& 61.66& 84.26&83.52\\ 
         PDANet~\cite{zhao2019pdanet}& 76.95& 68.05& -& 85.36&83.80\\\hline 
 \multicolumn{6}{c}{\textbf{Zero-shot VLMs}}\\  
         InstructBLIP~\cite{NEURIPS2023_9a6a435e}& 19.14& 37.17& 27.60& 42.86& 55.56\\ 
         Qwen2-VL-7b~\cite{wang2024qwen2}& 39.45& 53.54& 38.55& 68.40&69.97\\  
         LLaVA-Next~\cite{liu2024improved}& 54.93& 60.17& 53.36& 74.67&78.02\\ \hline 
 \multicolumn{6}{c}{\textbf{Instruction-tuned VLM}}\\  
         EmoVIT~\cite{xie2024emovit}& \textbf{83.36}& 68.09& 57.81& -&-\\ \hline  
         \rowcolor{gray!15} \textbf{EmoVLM-KD}& 79.83 & \textbf{79.51}& \textbf{73.91}&\textbf{88.90}&\textbf{89.59}\\ \hline
    \end{tabularx}
    \caption{Evaluation results on five benchmark VEA datasets (\%).}
    \label{tab:results}
\end{table*}

\subsection{Baseline Models}
\begin{itemize}
    \item \textbf{Conventional Vision Models.} We experimented with three widely used models for vision tasks: VGG16~\cite{simonyan2015very}, ResNet50~\cite{he2016deep}, and ViT~\cite{dosovitskiy2020image}. VGG16 is a deep convolutional neural network that was widely adopted in its early stages owing to its simple yet effective architecture. ResNet50, an improved deep residual network, introduces skip connections to address the vanishing gradient problem, thereby allowing for the training of significantly deeper networks. In addition to these CNN-based models, we also utilize Vision Transformer (ViT), which processes images as sequences of patches and leverages self-attention mechanisms to capture long-range dependencies, making it highly effective in various vision tasks.
    
    \item \textbf{VEA-specific Models.} We also compared models specifically designed for VEA with domain-specific training before the emergence of VLMs. WSCNet~\cite{yang2018weakly} is a weakly supervised network that enhances the performance by generating sentiment-specific soft maps to focus on emotionally relevant image regions. The Stimuli-Aware VEA model~\cite{yang2021stimuli} selects emotional stimuli and extracts features using the stimuli-organism-response (S-O-R) framework. MDAN~\cite{xu2022mdan} is a multilevel dependent attention network that improves performance by leveraging both global and local learning to bridge the affective gap in VEA. PDANet~\cite{zhao2019pdanet} is a deep attention network that improves the performance by enforcing polarity consistency through spatial and channel-wise attention.
    
    \item \textbf{Zero-shot VLMs.} To examine the impact of instruction tuning, we compared our approach with pre-trained VLMs under zero-shot conditions. InstructBLIP~\cite{NEURIPS2023_9a6a435e} is a vision-language model that leverages instruction tuning on 26 diverse datasets and achieves a state-of-the-art zero-shot performance across 13 datasets. LLaVA-Next~\cite{liu2024improved} is a large multimodal model that enhances visual reasoning and OCR performance by increasing the input resolution and improving visual instruction tuning. Qwen2-VL-7b~\cite{wang2024qwen2} is a VLM that offers advanced capabilities in visual understanding and multilingual text recognition. 
    
    \item \textbf{Instruction-tuned VLM.} We also evaluated our method against EmoVIT~\cite{xie2024emovit}, an instruction-tuned VLM specifically fine-tuned for emotion-recognition tasks. EmoVIT utilizes emotion-specific instructions, allowing the model to better understand and classify emotions in images and texts by closely aligning visual and emotional cues.

\end{itemize}

\subsection{Implementation Details}
In this study, we utilized the Qwen2-VL-7b model as a VLM, which was pre-trained and demonstrated a high performance across various VLM tasks. The model was fine-tuned on the emotion-specific instruction dataset using the AdamW optimizer, with a learning rate of 1e-4.

For the knowledge distillation stage, we employed ViT as the teacher model, which is known for its superior performance compared with conventional vision models. The ViT model was fine-tuned for each dataset to create domain-specific models for emotion prediction, enabling it to serve effectively as a teacher model. The student model, implemented as the distillation module, consisted of a linear layer and was trained with a learning rate of 1e-3. During this phase, only the parameters of the distillation module were updated, whereas the rest of the model remained frozen. The gate module, which integrates the outputs from the LLM and distillation module, was trained with a learning rate of 1e-3.

\subsection{Results}
We report the evaluation results based on accuracy in Table~\ref{tab:results}. As a result, the proposed EmoVLM-KD consistently outperformed conventional vision models, VEA-specific models, and zero-shot VLMs across multiple datasets, demonstrating its effectiveness in emotion recognition. ViT achieved 68.86\% on FI and 61.95\% on Emotion6, while our method showed higher accuracies of 79.51\% and 73.91\%, respectively. These results indicate that integrating VLMs with a distillation-based approach enhances the ability to capture emotion-related features more effectively than conventional vision models.

Although VEA-specific models are designed for affective computing, our approach achieved improved results. On the FI dataset, our model recorded 79.51\%, surpassing MDAN of 76.42\%. Similarly, on the Emotion6 dataset, our method reached 73.91\%, while the highest accuracy among VEA-specific models was 61.66\%. 

Zero-shot VLMs, which are not specifically adapted for emotion recognition, showed lower performance in this task. LLaVA-Next achieved 60.17\% on FI and 53.36\% on Emotion6, which are lower than the performance of our method. This result highlights the effectiveness of leveraging VLMs in combination with a distillation module for improving emotion understanding in VLMs.

Furthermore, on the Flickr and Instagram datasets, where each image is assigned one of two emotion labels, our model achieved 88.90\% and 89.59\% , respectively. These results indicate that our approach not only excels in recognizing fine-grained emotional distinctions but effectively adapts to binary classification tasks, further underscoring its robustness in diverse emotion recognition scenarios.

While our model did not achieve the highest accuracy on Emoset, it remained competitive. More importantly, its consistent superiority across FI, Emotion6, Flickr, and Instagram underscores its robustness and generalizability. 

\subsection{Ablation Study}
\subsubsection{Effect of the Knowledge Distillation}
To evaluate the effectiveness of integrating a distilled module into a VLM, we conducted an experiment using the FI dataset. We compared the number of trainable parameters and accuracy across different module configurations: (1) an instruction-tuned VLM, (2) a distillation module, (3) a vision model used as the teacher model for distillation (i.e., ViT), and (4) our proposed EmoVLM-KD model, which combines the VLM with the distillation module. As shown in Table~\ref{tab:ablation_1}, VLM achieved an accuracy of 71.19\%, whereas EmoVLM-KD improved to 79.51\%, indicating an approximately 8\% performance increase. Notably, this enhancement was achieved with only a 0.3K increase in trainable parameters, which constitutes only 0.003\% of the VLM’s trainable parameters. 

Another notable observation was that the distilled module outperformed the teacher model ViT. While ViT achieved an accuracy of 68.86\%, that of the distilled module reached 79.51\%. This suggests that during the knowledge distillation process, the distilled module not only learns to replicate the teacher model's prediction patterns, but also benefits from direct supervision with ground-truth labels, leading to its superior performance.

\begin{table}
    \centering
    \begin{tabular}{l|cc}
    \hline
         &  \textbf{Parameter count}& \textbf{Accuracy}\\ \hline
         (1) VLM&  1,091,870,720& 71.19\\
         (2) Distill&  3,679,240& 77.37\\
         (3) ViT&  85,804,808& 68.86\\
         (4) EmoVLM-KD&  1,095,550,096& \textbf{79.51}\\ \hline
    \end{tabular}
    \caption{The number of trainable parameters and accuracy (\%) with different module configurations.}
    \label{tab:ablation_1}
\end{table}

\subsubsection{Depth Analysis}
\begin{table}[h]
    \centering
    \resizebox{0.48\textwidth}{!}{%
    \begin{tabular}{l|cc|cc}
        \hline
        \multirow{2}{*}{\textbf{Layer configuration}} & \multicolumn{2}{c|}{\textbf{Parameter count}} & \multicolumn{2}{c}{\textbf{Accuracy}} \\
        \cline{2-5}
         & \textbf{FI} & \textbf{Emotion6} & \textbf{FI} & \textbf{Emotion6} \\
        \hline
        {[1024]}                         & 3,679,240& 3,677,190& \textbf{79.51}& \textbf{73.91}\\
        {[1024, 512]}                       & 4,199,944& 4,198,918
& 73.98& 69.02\\
        {[1024, 512, 256]}                  & 4,329,224& 4,328,710& 73.80& 68.86\\
        {[1024, 512, 256, 128]}             & 4,361,096& 4,360,838& 72.43& 67.68\\
        {[1024, 512, 256, 128, 64]}         & 4,368,840& 4,368,710& 68.59& 63.30\\
        \hline
    \end{tabular}}
    \caption{The number of trainable parameters and accuracy (\%) based on different layer configurations. In each layer configuration, the numbers inside the brackets represent the hidden dimensions, and the count of these numbers indicates the number of layers.}
    \label{tab:ablation_2}
\end{table}
In this study, we aimed to determine the most suitable layer size for knowledge distillation by examining how small the layers could be in terms of parameter count without compromising the performance. We evaluated the performance based on the accuracy using the FI and Emotion6 datasets by conducting a series of experiments with simple linear layers stacked from one to five layers after the visual encoder of the VLM. As shown in Table~\ref{tab:ablation_2}, the results showed that both datasets achieved the best performance with shallower layers despite having fewer parameters. For instance, the FI dataset achieved an accuracy of 79.51\% with a single hidden layer of 1024 dimensions, whereas the Emotion6 dataset achieved an accuracy of 73.91\% with the same hidden layer, thus outperforming deeper architectures.

\subsubsection{Hyperparameter Analysis}\label{sec:ablation_3}
\begin{figure}[ht]
    \centering
    \includegraphics[width=0.5\textwidth]{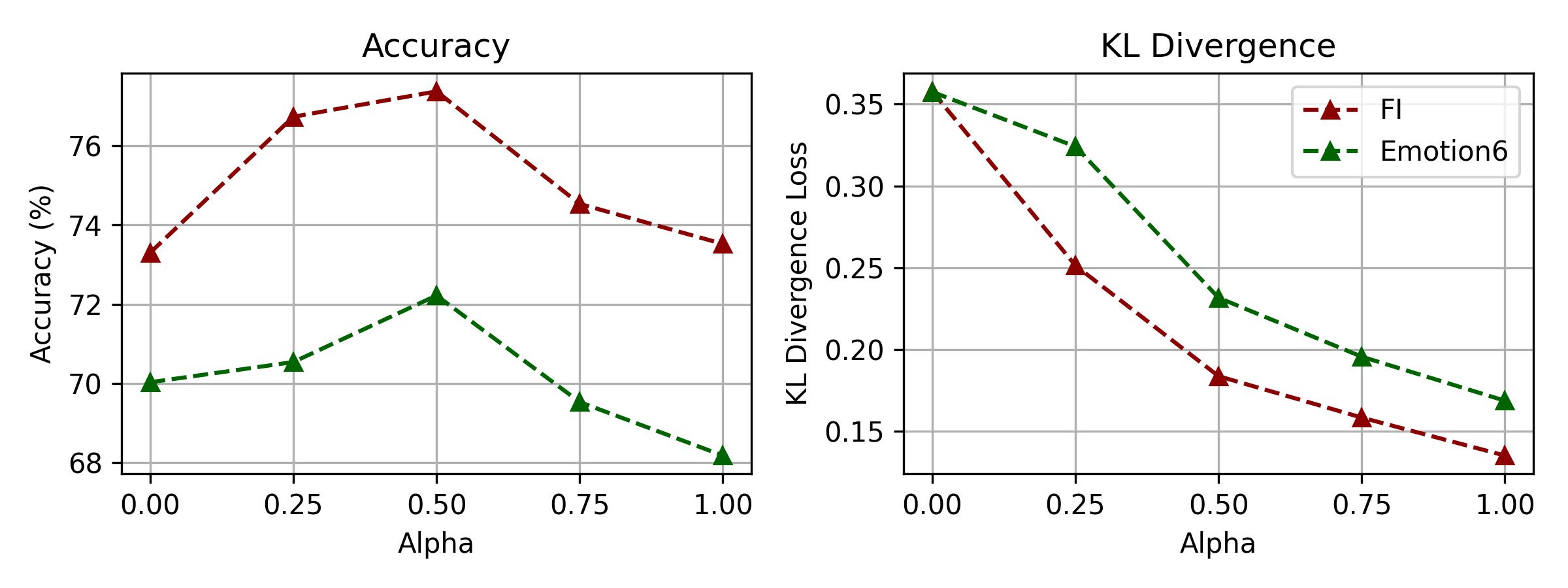}
    \caption{Accuracy (\%) and KL divergence loss for different alpha hyperparameter settings.} 
    \label{fig:ablation_3}
\end{figure} 

This study aimed to examine the effect of different ratios between the two loss functions in the knowledge distillation process and to determine the optimal alpha value that balances these losses for the best performance. We used the FI and Emotion6 datasets to evaluate the accuracy of assessing the predictive performance and KL divergence to measure the difference between the prediction pattern of the teacher model and that of the student model. As depicted in Figure~\ref{fig:ablation_3}, the results showed that both datasets achieved the highest accuracy when the alpha value was set to 0.5. As expected, a higher alpha value had a greater influence on KL divergence, thereby reducing KL loss. We concluded that an alpha value of 0.5 represents the best trade-off between KL loss and accuracy.

\subsubsection{Gate Architecture} \label{sec:ablation_4}

\begin{table}[b]
    \centering  
    \begin{tabular}{l|c} \hline
         &  \textbf{Accuracy} \\ \hline
         (1) Concat \& Linear&  79.51\\
         (2) MoE&  79.24\\
         (3) Bilinear&  78.56\\
         (4) DynamicWeighting&  78.14\\
         (5) CrossGating&  78.01\\ \hline
    \end{tabular}
    \caption{Accuracy (\%) across the five gating methods.}
    \label{tab:ablation_4}
\end{table}

\begin{figure*}[t]
    \centering
    \includegraphics[width=\textwidth]{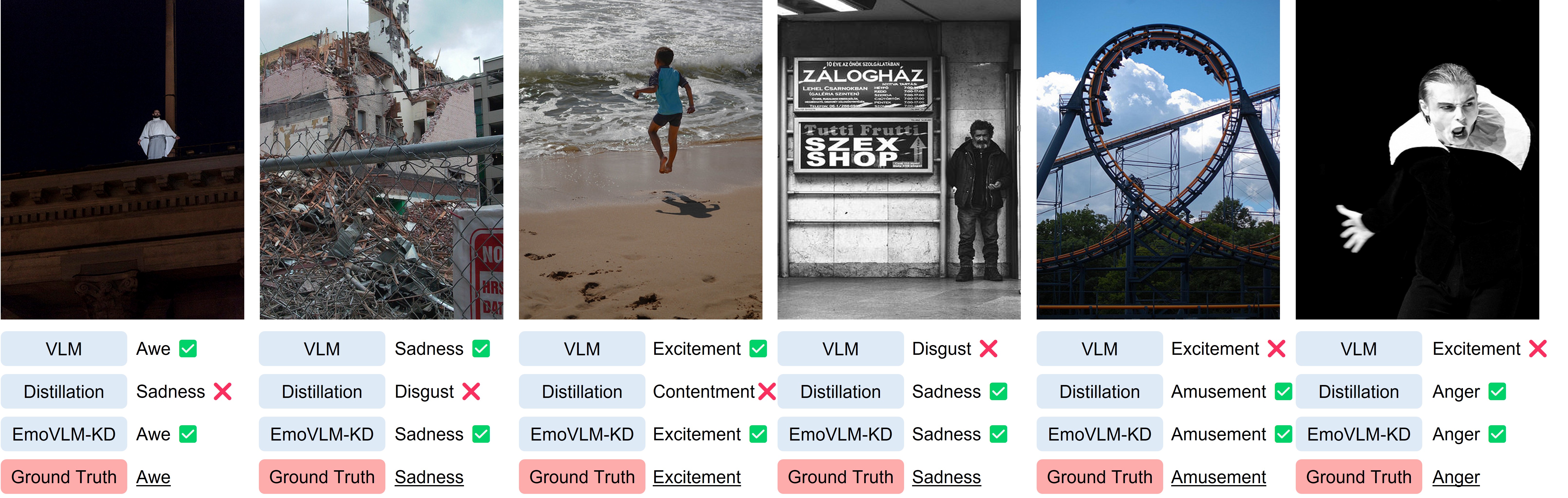}
    \caption{Qualitative examples of EmoVLM-KD's prediction.} 
    \label{fig:qualitative}
\end{figure*} 

The purpose of this study was to explore how to effectively combine predictions from the VLM and distillation modules to enhance the prediction accuracy. 

We used the FI dataset to measure accuracy and compared five different gating methods: (1) Concat \& Linear, which concatenates the two predictions into a single vector and passes them through a linear layer to produce the final output; (2) MoE (i.e., mixture of experts), which consists of multiple expert networks with a softmax function to assign weights to each expert, followed by a linear layer for the final result; (3) bilinear pooling gate, which applies a bilinear operation of the form $v_1^T \cdot W \cdot v_2$ to retain an output vector, followed by a linear transformation. Here, 
$v_1$ and $v_2$ represent the outputs of the VLM and distillation modules, respectively. (4) The dynamic weighting gate computes weights $w1$ and $w2$ through a softmax function applied to a linear transformation of the concatenated predictions, and the final output is $w_1 \cdot v_1 + w_2 \cdot v_2$. (5) CrossGating, which calculates gates $g1$ and $g2$ using sigmoid functions on linearly transformed opposite predictions, and combines them as $g_1 \cdot v_1 + g_2 \cdot v_2$. The results in Table ~\ref{tab:ablation_4} indicate that the performance differences among these methods were minimal, but the (1) Concat \& Linear approach achieved the highest accuracy of 79.51\%.

\subsection{Qualitative Examples}
Figure~\ref{fig:qualitative} presents the predictions of the VLM and distillation module within EmoVLM-KD, along with the final prediction of EmoVLM-KD and the ground truth. The three examples on the left are cases where the VLM correctly predicts the emotion category, while the distillation module fails. Conversely, the three examples on the right are cases where the VLM makes incorrect predictions, whereas the distillation module correctly predicts the emotion category. In both cases, we observed that EmoVLM-KD selects the correct prediction from either the VLM or distillation module, aligned with the ground truth. This demonstrates that EmoVLM-KD effectively combines the predictions of both models to achieve accurate predictions. 

For example, in the leftmost example, given an image with the ground truth label ``awe'', the VLM predicts ``awe'', while the distillation module predicts ``sadness'', showing different predictions. However, EmoVLM-KD correctly predicts ``awe'', demonstrating its ability to effectively integrate the outputs of both modules for more accurate emotion recognition. 

On the other hand, in the fourth example, the VLM predicts ``disgust'', while the distillation module predicts ``sadness'', with the ground truth label being ``sadness''. In this case, only the distillation module provides the correct prediction. EmoVLM-KD assigns a higher weight to the distillation module's prediction, thereby leading to an accurate final prediction. This suggests that EmoVLM-KD adaptively adjusts the contribution of each model to improve overall predictive accuracy. 

\section{Conclusion}
In this study, we present EmoVLM-KD, a novel approach to enhance VEA by transferring the capabilities of a vision model to an instruction-tuned VLM. We introduce a lightweight module to the visual encoder and employ a gate module to integrate the outputs from the VLM and distillation module. This approach achieves superior performance on multiple VEA benchmark datasets with minimal additional parameters. The experimental results demonstrate that EmoVLM-KD surpasses not only domain-specific models, but also a diverse range of VLM-based approaches, highlighting the effectiveness of our knowledge distillation and integration strategies. This approach suggests a promising direction for further improving the domain-specific VLM performance.

Future work will focus on extending the VEA capabilities of EmoVLM-KD to predict a wider range of emotion categories and incorporate additional modalities, paving the way for more comprehensive emotion recognition systems.

\section*{Acknowledgment} 
This work was supported in part by the Culture, Sports and Tourism R\&D Program through the Korea Creative Content Agency grant funded by the Ministry of Culture, Sports and Tourism in 2023 (Project Name: Development of AI-based Interactive Multi-Modal Storytelling 3D Scene Creation Technology for User Engagement, Project Number: RS-2023-00227917, Contribution Rate: 50\%), and in part by the MSIT (Ministry of Science and ICT), Korea, under the ICAN (ICT Challenge and Advanced Network of HRD) support program (RS-2024-00436934), and the ITRC (Information Technology Research Center) program (RS-2024-00436936) supervised by the IITP (Institute for Information \& Communications Technology Planning \& Evaluation).